\documentclass[aps,pre,twocolumn,10pt,showpacs]{revtex4-1}

\usepackage[latin1]{inputenc}
\usepackage{amsmath}
\usepackage{amsfonts}
\usepackage{amssymb}
\usepackage{graphicx,hyperref}
\usepackage{amssymb}
\usepackage{mathptmx}

\begin{document}

\title{Wakefield generation in magnetized plasmas}

\author{Amol Holkundkar} 
\email[E-mail address: ]{amol.holkundkar@physics.umu.se}
\affiliation{Department of Physics, Ume{\aa} University, SE--901 87 Ume{\aa}, Sweden}

\author{Gert Brodin} 
\email[E-mail address: ]{gert.brodin@physics.umu.se}
\affiliation{Department of Physics, Ume{\aa} University, SE--901 87 Ume{\aa}, Sweden}

\author{Mattias Marklund}
\email[E-mail address: ]{mattias.marklund@physics.umu.se}
\affiliation{Department of Physics, Ume{\aa} University, SE--901 87 Ume{\aa}, Sweden}

\begin{abstract}
We consider wakefield generation in plasmas by electromagnetic 
pulses propagating perpendicular to a strong magnetic field, in the 
regime where the electron cyclotron frequency is equal to or  larger 
than the plasma frequency. PIC-simulations reveal that for moderate 
magnetic field strengths previous results are re-produced, and the 
wakefield wavenumber spectrum has a clear peak at the inverse skin 
depth. However, when the cyclotron frequency is significantly larger 
than the plasma frequency, the wakefield spectrum becomes broad-band, 
and simultaneously the loss rate of the driving pulse is much enhanced. 
A set of equations for the scalar and vector potentials reproducing 
these results are derived, using only the assumption of a weakly 
nonlinear interaction.
\end{abstract}

\maketitle

\section{Introduction}

Since the pioneering work in Ref. \cite{Tajima-Dawson}, where the excitation
of a plasma oscillations by a short laser pulse was considered, much
interest has been devoted to wakefield generation in plasmas. To a large
extent this has been due to the wakefield acceleration concept (see, e.g., \cite%
{Tajima-Dawson,Kirsanov-1987,Phys-reports,Faure-etal,Nature-review,Martins-etal}), where the longitudinal field
traveling close to the speed of light in vacuum is used as an effective
source for particle acceleration. However, the wakefield properties has also
been study from a more theoretical point of view 
\cite{Mironov1990,Chen-Alfven,Multi-wake,Wadhwani2002,Shukla-Hybrid,PRE1998}. Most studies of
wakefield generation has been done for unmagnetized plasmas. This is
justified by the fact that in most experiments $\omega _{c}/\omega _{p}\ll 1$%
, where $\omega _{c}=qB_{0}/m$ is the electron cyclotron frequency, $\omega
_{p}=(n_{0}q^{2}/\varepsilon _{0}m)^{1/2}\,$is the plasma
frequency\thinspace \thinspace\ $B_{0}$ is the unperturbed magnetic field
strength, \thinspace $n_{0}$ is the unperturbed electron density, and $q$
and $m$ is the electron charge and mass respectively. As can be expected,
the wakefield properties is not much affected by the magnetic field in this
regime. However, for $\omega _{c}/\omega _{p}\gtrsim 1$, and for the
exciting source propagating non-parallel to the the external magnetic
field, the wakefield is significantly altered by the non-zero $B_{0}$\cite%
{Wadhwani2002,Shukla-Hybrid,PRE1998,Yooshii1997}. In particular, as found by
e.g. Refs. \cite{Yooshii1997,PRE1998}, for the EM-pulse traveling
perpendicular to the magnetic field, the excited mode becomes an
extra-ordinary mode, with a significant electromagnetic part, and a non-zero
group velocity.

In the present paper we will re-consider the excitation of the
extra-ordinary mode wakefield in a strongly magnetized plasma, allowing for 
$\omega _{c}/\omega _{p}\gg 1$. In this regime the group-velocity of the wakefield 
may approach the speed of light in vacuum $c$. In then turns out that
the excitation process is significantly altered, as compared to the regime of
a more modest value of $B_{0}$. From Particle-In-Cell (PIC) simulations we find that when 
$B_{0}$ is increased, the wakefield changes from an approximately
monochromatic field at low $\omega _{c}/\omega _{p}$, to a broadband one at $%
\omega _{c}/\omega _{p}\gg 1$. At the same time the energy loss rate of the
exciting high-frequency ordinary mode is much increased with 
increasing $\omega_{c}/\omega _{p}$. A reduced system of equations is derived,
generalizing previous Eqs. \cite{PRE1998} avoiding the division into fast
and slow time scales. The reduced equations are solved numerically, and a
good agreement with the full 1D PIC simulations is found. 

\section{PIC-simulations of wake field generation in a magnetized plasma}

Here we will study an ordinary mode propagating parallel to an external
magnetic field, $\mathbf{B}_{0}=B_{0}\mathbf{x}$. In the absence of the
magnetic field, it is well known that the ponderomotive force due to a short
electromagnetic pulse will excite a wakefield of electrostatic oscillations 
\cite{Tajima-Dawson,Kirsanov-1987}. Here we will focus on the effects due to
the external magnetic field \cite{Yooshii1997,PRE1998}. We note that this is
more likely to be relevant for experiments in the microwave regime \cite%
{Parchamy2005}, rather than the regime of optical lasers.

\begin{figure}[h]
\centering \includegraphics[width=.9\columnwidth,trim=0cm 4cm 0cm 0cm]{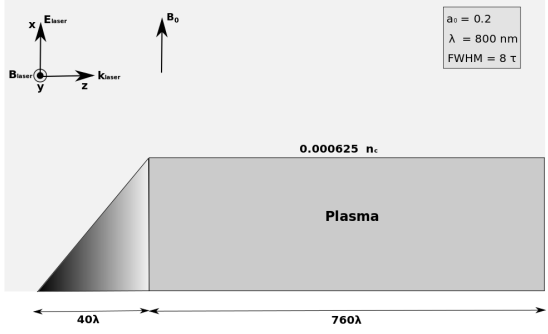}
\caption{Simulation geometry for the study.}
\label{geometry}
\end{figure}

The 1D Particle-In-Cell simulation (LPIC++)\cite{lpic} is carried out to study the
effect of external magnetic field on the wakefield generation in magnetized
plasma. The typical simulation geometry is shown in Fig. 1, where the
exciting electromagnetic (EM) high-frequency pulse is taken to be an ordinary
mode. For all the results presented here the EM field amplitude of the
ordinary mode driver is considered as 0.2 (in dimensionless unit 
$a_{0}=eE/m\omega c$, with $\omega $ being the frequency), space and time
are taken in units of laser wavelength ($\lambda $) and one laser cycle $%
\tau =\lambda /c$ respectively. The 800 nm laser with FWHM duration of 
8$\tau $ and an initially Gaussian shape propagates through the plasma along
the $z$-direction with electric field along $x$ and the laser magnetic
field along the $y$-direction \cite{NOTE}. We have modified the code in order to extend its
ability to incorporate the presence of an external magnetic field. The external
magnetic field is taken to be along the $x$-direction. The plasma of length 800
$\lambda $ is considered with a 40$\lambda $ of the ramp region at the front.
In the ramp region the plasma density increases linearly from 0 to 0.000625 $%
n_{c}$ (Fig. \ref{geometry}), where $n_{c}=\varepsilon_{0} m \omega^2/q^2$.

\begin{figure}[h]
\centering \includegraphics[width=.7\columnwidth,angle=270]{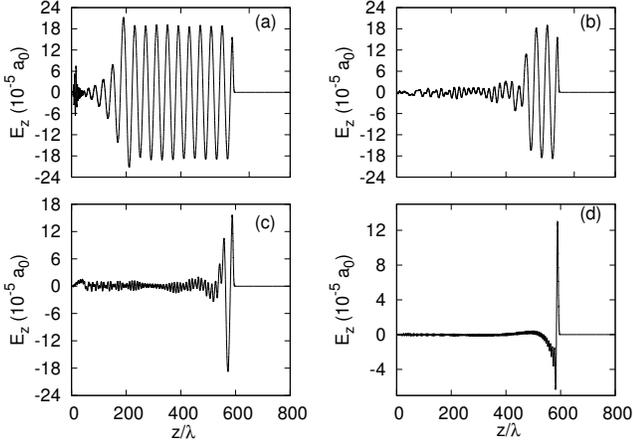}
\caption{Spatial profiles of the longitudinal wakefield at 600$\protect\tau$
in presence of external magnetic field with ratio $\protect\omega_c/\protect%
\omega_p =$ 0.5 (a), 2 (b), 4 (c) and 8 (d) respectively are presented using
PIC simulation. }
\label{WF_Ex_PIC}
\end{figure}

\begin{figure}[h]
\centering \includegraphics[width=.7\columnwidth,angle=270]{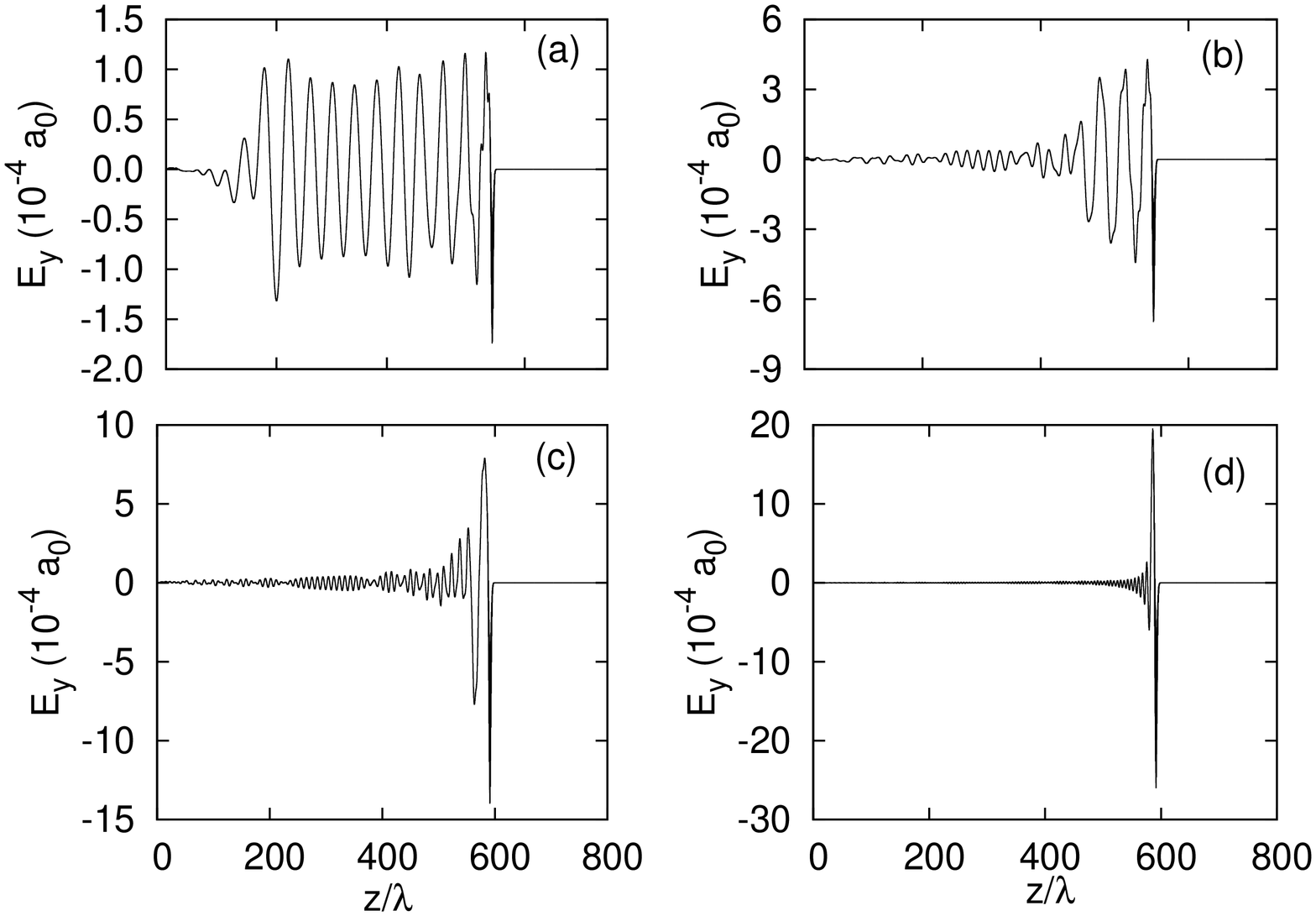}
\caption{Spatial profiles of the transverse wakefield at 600$\protect\tau$ in
presence of external magnetic field with ratio $\protect\omega_c/\protect%
\omega_p =$ 0.5 (a), 2 (b), 4 (c) and 8 (d) respectively are presented using
PIC simulation. }
\label{WF_Ez_PIC}
\end{figure}

Due to the external magnetic field the wakefield gets both a longitudinal
and a transverse part. In Fig. \ref{WF_Ex_PIC} the longitudinal part of the electric field
wakefield is shown when varying the value of the external magnetic field.
Similarly the transverse component is shown in Fig. \ref{WF_Ez_PIC}.

A number of features from the PIC-simulations agree with the model and
results presented in Ref. \cite{PRE1998} (see also the discussion and
simulation results in Ref. \cite{Yooshii1997})

\begin{enumerate}
\item The ratio of the electromagnetic to electrostatic field field
amplitude scales as $\omega _{c}/\omega _{p}$ when $B_0$ is varied. 
This is in accordance with the wake field being an extra-ordinary wave mode.

\item By observing the position of the drop in the wake field amplitude, we
can deduce that wake field propagates with a group velocity $v_{g\mathrm{wf}%
}\approx c\omega _{c}^{2}/(\omega _{p}^{2}+\omega _{c}^{2})$, which is in
accordance with Refs. \cite{Yooshii1997,PRE1998}
\end{enumerate}

However, there are also some properties of the PIC-simulations that contain
new results, as compared to previous findings.

\begin{figure}[ht]
\centering \includegraphics[width=.7\columnwidth,angle=270]{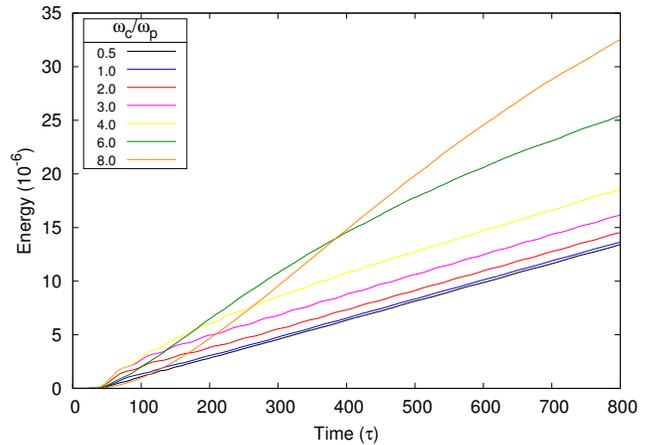}
\caption{Temporal evolution of the total energy of the wakefield (ES and EM
components) for different values of the external magnetic fields using PIC
simulation.}
\label{loss_PIC}
\end{figure}

\begin{enumerate}
\item The wakefield does have a small but nonzero trailing part at a
distance more than $\left( v_{g}-v_{g\mathrm{wf}}\right) T$, where $%
v_{g}\approx c$ is the group velocity of the ordinary mode, and $T$ is the
time after the entrance of the driving pulse. The relative energy content 
of the wakefield in this tail is larger for larger value of $\omega _{c}/\omega _{p}$. 

\item The approximation that the evolution of the wakefield is slow in a
frame moving with the pulse is not entirely accurate. Furthermore, the
theory of Ref. \cite{PRE1998} becomes increasingly inadequate when the
magnetic field strength is increased such that $\omega _{c}/\omega _{p}\gg 1$. 
In particular, Ref. \cite{PRE1998} predicts that the loss-rate of the
driving pulse is approximately independent of $\omega _{c}/\omega _{p}$. In
Fig. \ref{loss_PIC} the total energy $\propto S_z^{WF} + |E_z^2|$, where $S_z^{WF} = |E_x B_y - E_x B_y - (B_x-B_0) E_y |$ is the EM part, contained in the wakefield is plotted as a function of time
for different values of $\omega _{c}/\omega _{p}$.  We see that the loss rate of 
the driving pulse due to the generated wakefield is approximately constant 
until $\omega _{c}/\omega _{p}\gtrsim 2$, but for larger values of this ratio it increases 
rapidly.
\end{enumerate}

One way to understand the deviance from the models in Ref. \cite{PRE1998}
is to investigate the validity of the basic assumptions. A key ingredient in
the derivation is that the evolution should be slow in a frame moving with
the group velocity of the driver. A consequence of this is that the phase
velocity of the wakefield should match the group velocity of the driver.
Since the wakefield is dispersive, such a condition determines the
dominating frequency and wavenumber to be excited. Noting that the plasma is underdense with respect
to the driver frequency, the group velocity of the driver is close to $c$,
which matches the wakefield phase velocity in case the excited wakefield frequency 
is $\omega _{p}$ and wavenumber $\omega _{p}/c$. Such a condition can be seen to hold for the leading
part of the wakefield, but for a strong static field this part actually
makes up for a small part of the energy content of the excited
extra-ordinary mode. This is illustrated in Fig. \ref{FFT_Ez}, where the frequency
spectrum of the wakefield is computed as a function of the static magnetic
field for a fixed position and time of interaction. For $\omega _{c}/\omega _{p}\gtrsim 2$
we see that the spectrum ceases to peak at $\omega _{p}$, and as a
consequence the assumption that the evolution of the wakefield is slow in a
system moving with the group velocity cannot be accurate. In the rest of the
paper the aim is to investigate the physics of wakefield generation in the
regime where the basic equations of Ref. \cite{PRE1998} needs to be improved.

\section{The weakly relativistic 1-D model}
For 1D-perturbations (with variations along $z$), the cold relativistic
fluid equations can be written. 
\begin{equation}
\frac{\partial n}{\partial t}+\frac{\partial }{\partial z}(nv_{z})=0
\label{continuity}
\end{equation}%
and 
\begin{equation}
\left( \frac{\partial }{\partial t}+v_{z}\frac{\partial }{\partial z}\right)
\left( \gamma \mathbf{v-}\frac{q}{m}\mathbf{A}\right) \mathbf{=}\frac{q}{m}%
\frac{\partial \Phi }{\partial z}\mathbf{z}+\omega _{c}\mathbf{v\times x}
\label{momentum}
\end{equation}%
where we have assumed the field to be of the form\textbf{\ }$\mathbf{E}%
=-\partial \mathbf{A/}\partial t-\nabla \Phi $, $\mathbf{B}=\nabla \times 
\mathbf{A+B}_{0}$, the constant magnetic field is $\mathbf{B}_{0}=B_{0}%
\mathbf{x}$\thinspace , $\gamma =(1-v^{2}/c^{2})^{1/2}$ is the relativistic
factor, and $\omega _{c}=qB_{0}/m$ is the (constant) cyclotron frequency. For
the case where the vector potential was a weakly modulated function, $%
\mathbf{A=}\widetilde{\mathbf{A}}(z,t)\exp (ikz-\omega t)+\mathrm{c.c.}$,
where c.c. stands for complex conjugate, Ref. \cite{PRE1998} derived a
coupled set for $A$ (with $\widetilde{\mathbf{A}}\mathbf{=}A\mathbf{x}$) and 
$\Phi $, where $A$ was the vector potential amplitude for the ordinary mode
of high-frequency ($\omega \gg \omega _{p}$), and $\Phi $ described the
electrostatic potential of the extra-ordinary mode. It was also noted in
Ref. \cite{PRE1998} that the wake-field was also found to have en
electromagnetic part, that could be computed from $\Phi $. The main
modification of the wake-field generation due to the static magnetic field $%
\mathbf{B}_{0}$ was to introduce an electromagnetic component of the wake
field, which in turn resulted in a non-zero group velocity. The
PIC-simulations of the previous section confirms that these properties are
essential ingredients when the external magnetic field is introduced.
However, when the ratio $\omega _{c}/\omega _{p}$ is increased beyond unity,
we also saw that additional features were introduced. In particular, the
assumption of Ref. \cite{PRE1998} that the fields evolve slowly in a frame moving with the group
velocity, becomes less accurate with increasing value of $\omega _{c}/\omega
_{p}$. In the rest of this section we will derive the coupled equations for
an extra-ordinary mode driven by an ordinary mode, without any additional
assumptions than those of an 1D geometry and a weak nonlinearity. In
particular, we will avoid making a WKB-ansatz for the high-frequency
ordinary mode. We use the Coulomb gauge and write the vector potential as $%
\mathbf{A=}A\mathbf{x+}A_{w}\mathbf{y}$. Thus the component $A_{w}$ is
associated with the electromagnetic component of the nonlinearly driven
extra-ordinary mode, and $\Phi $ with its electrostatic component. The
fields of the ordinary mode is derived solely from $A$.

From the $x$-component of the momentum equation (\ref{momentum}) we obtain 
\begin{equation}
v_{x}=-\frac{qA}{mc}\left( 1-\frac{q^{2}A^{2}}{2m^{2}c^{2}}\right)
\label{momentum-x}
\end{equation}%
which we substitute into Ampere's law, to give 
\begin{equation}
\frac{\partial ^{2}A}{\partial t^{2}}-c^{2}\frac{\partial ^{2}A}{\partial
z^{2}}+\omega _{p}^{2}A+\frac{q^{2}}{\varepsilon _{0}m}\delta nA-\frac{%
\omega _{p}^{2}q^{2}}{2m^{2}c^{4}}A^{3}=0  \label{New1}
\end{equation}%
where $\omega _{p}=(n_{0}q^{2}/\varepsilon _{0}m)^{1/2}$ is the plasma
frequency associated with the unperturbed electron density $n_{0}$. Next we
study the z-component of the momentum equation. We keep variables that are
linear in the extra-ordinary wave-mode, and include quadratic nonlinearities
in the ordinary wave mode variables. Essentially this nonlinearity turns out to be the
ponderomotive force, but keeping also the second-harmonic term in addition
to the low-frequency contribution. The z-component of the velocity is then
expressed in terms of the electrostatic potential (through the z-component
of Ampere's law) and the ordinary variables are expressed in terms of $A$.
Finally the y-component of the velocity is expressed in terms of $A_{w}$
(from the $y$-component of Ampere's law). The equation is then expressed
solely in terms of the potentials, and we obtain
\begin{equation}
\left( \frac{\partial ^{2}}{\partial t^{2}}+\omega _{p}^{2}\right) \frac{%
\partial \Phi }{\partial z}+\frac{\omega _{c}}{c}\left[ \frac{\partial ^{2}}{%
\partial t^{2}}-c^{2}\frac{\partial ^{2}}{\partial z^{2}}\right] A_{w}=-%
\frac{\omega _{p}^{2}q}{2mc^{2}}\frac{\partial A^{2}}{\partial z}.
\label{New2}
\end{equation}%
Finally, we study the $y$-component of the momentum equation, and substitute
the velocities in terms of the potentials in the same way as previously,
which leads to 
\begin{equation}
\left[ \frac{\partial ^{2}}{\partial t^{2}}-c^{2}\frac{\partial ^{2}}{%
\partial z^{2}}+\omega _{p}^{2}\right] A_{w}=c\omega _{c}\frac{\partial \Phi 
}{\partial z}  \label{New3}
\end{equation}%
By using Poisson's eq. the density perturbation can be expressed in terms of
the potential, $\partial ^{2}\Phi /\partial z^{2}=q\delta n/\varepsilon _{0}$%
, and Eqs. (\ref{New1})-(\ref{New3}) forms a closed set for $A,A_{w}$ and $%
\Phi $. For zero external magnetic field $\omega _{c}\rightarrow 0$, the
coupling to the electromagnetic part of the wake-field $A_{w}$ vanishes, and
(\ref{New1}) together with (\ref{New2}) form a closed set for $A$ and $\Phi $%
. By eliminating $\partial \Phi /\partial z$ in terms of $A_{w}$ using (\ref%
{New3}) in Eqs.(\ref{New1}) and (\ref{New2}) we may get a coupled set of two
equations for $A$ and $A_{w}$, which however has fourth order derivatives
from the wave operator of the extra-ordinary mode. Specifically, we obtain 
\begin{eqnarray}
&&
\left[ \left( \frac{\partial ^{2}}{\partial t^{2}}+\omega _{p}^{2}\right)
\left( \frac{\partial ^{2}}{\partial t^{2}}-c^{2}\frac{\partial ^{2}}{%
\partial z^{2}}+\omega _{p}^{2}\right) \right.
\nonumber \\ &&\qquad 
\left.
+\omega _{c}^{2}\left( \frac{\partial
^{2}}{\partial t^{2}}-c^{2}\frac{\partial ^{2}}{\partial z^{2}}\right) %
\right] A_{w}=-\frac{\omega _{p}^{2}\omega _{c}q}{2mc^{2}}\frac{\partial
A^{2}}{\partial z}  \label{New4}
\end{eqnarray}
which confirms that the electromagnetic part $A_{w}$ becomes proportional to
the external magnetic field through $\omega _{c}$.

In order to perform a numerical analysis, let us first put the weakly
relativistic 1D-model in dimensionless form. Introducing the normalized
variables $t_{n}=\omega _{p}t$, $z_{n}=z\omega _{p}/c$, $A_{n}=qA/mc$, $\Phi
_{n}=q\Phi /mc^{2}$, $A_{nw}=qA_{w}/mc$, we obtain
\begin{eqnarray}
\frac{\partial ^{2}A}{\partial t^{2}}-\frac{\partial ^{2}A}{\partial z^{2}}%
+A-\frac{\partial ^{2}\Phi }{\partial z^{2}}A-\frac{1}{2}A^{3} &=&0
\label{laser-evolution} \\[2mm]
\left( \frac{\partial ^{2}}{\partial t^{2}}+1\right) \frac{\partial \Phi }{%
\partial z}+\frac{\omega _{c}}{\omega _{p}}\left[ \frac{\partial ^{2}}{%
\partial t^{2}}-\frac{\partial ^{2}}{\partial z^{2}}\right] A_{w} &=&-\frac{1%
}{2}\frac{\partial A^{2}}{\partial z}  \label{wake-1} \\[2mm]
\left[ \frac{\partial ^{2}}{\partial t^{2}}-\frac{\partial ^{2}}{\partial
z^{2}}+1\right] A_{w} &=&\frac{\omega _{c}}{\omega _{p}}\frac{\partial \Phi 
}{\partial z}  \label{Wake-2}
\end{eqnarray}%
where we have left out the index $n$ denoting normalized variables for
notational convenience. It is straightforward to show that the above system
conserves total wave energy $\int Wdz$ if the energy flux across the
boundaries vanishes. The energy density $W=W_{\mathrm{d}}+W_{\mathrm{wf}}$
where the wake field contribution to the energy density is 
\begin{eqnarray}
&&
W_{\mathrm{wf}}=\frac{1}{2}\left[ \left( \frac{\partial \Phi }{\partial z}%
\right) ^{2}+\left( \frac{\partial A_{w}}{\partial z}\right) ^{2}+\left( 
\frac{\partial A_{w}}{\partial t}\right) ^{2}
\right.
\nonumber \\ && \qquad
\left.
+\left( \frac{\partial \Phi }{%
\partial z\partial t}\right) ^{2} +\left( \frac{\partial ^{2}A_{w}}{\partial
t^{2}}-\frac{\partial ^{2}A_{w}}{\partial z^{2}}\right) ^{2}\right]
\label{energy-density}
\end{eqnarray}
with the terms form left to right representing longitudinal electric field
energy density, magnetic field energy density, perpendicular electric
field energy density, longitudinal kinetic energy density, and perpendicular
kinetic energy density. Similarly, the energy density due to the driving
ordinary mode is 
\begin{equation}
W_{\mathrm{d}}=\frac{1}{2}\left[ \left( \frac{\partial A}{\partial t}\right)
^{2}+\left( \frac{\partial A}{\partial z}\right) ^{2}+A^{2}\left(1-\frac{\partial
^{2}\Phi }{\partial z^{2}}\right)-\frac{1}{2}A^{4}\right]
\label{energy-dens-driver}
\end{equation}%
where the terms from left to right represent electric field energy density,
magnetic field energy density, lowest order (linear) kinetic energy density,
density perturbation correction to kinetic energy density, and finally
relativistic correction to kinetic energy density. The energy conservation
law for Eqs. (\ref{laser-evolution})- (\ref{Wake-2}) together with the
expressions (\ref{energy-density}) and (\ref{energy-dens-driver}) will be
used in the next section for calculating the energy loss rate of the driving pulse
due to the wakefield generation.

\section{Numerical comparison}

Next we will compare the weakly relativistic 1D model with the results from
the PIC-simulations. The temporal evolution of the quantities $A(z,t)$, $%
A_{w}(z,t)$ and $\partial \Phi (z,t)/\partial z$ is studied by numerically
solving the dimensionless Eqs. (\ref{laser-evolution})- (\ref{Wake-2}) using
a standard explicit Leap-Frog scheme combined with the following initial and
boundary conditions,
\begin{equation*}
A_{w}(z,0)=0, \quad 
\partial \Phi (z,0)/\partial z=0, \quad
A(z,0)=0
\end{equation*}%
and 
\begin{equation*}
A(0,t)=A_{0}\cos(\omega t)\exp\left[ -4\left( \frac{t}{\tau _{\rm FWHM}}%
\right) ^{2}\right]
\end{equation*}%
Here $\tau _{\rm FWHM}$ is FWHM pulse duration of the laser which is considered
to be $\tau _{\rm FWHM}=1.25$ in dimensionless units and is equivalent to the
laser pulse duration used in PIC simulations.

\begin{figure}[ht]
\centering \includegraphics[width=.7\columnwidth,angle=270]{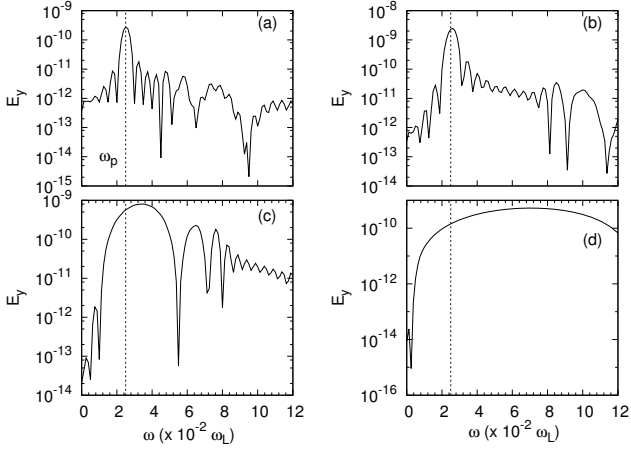}
\caption{Fourier spectrum of the transverse wakefield calculated at 595$%
\protect\lambda$ in presence of external magnetic field with ratio $\protect%
\omega_c/\protect\omega_p =$ 0.5 (a), 2 (b), 4 (c) and 8 (d) respectively
are presented using PIC simulation. }
\label{FFT_Ez}
\end{figure}

\begin{figure}[ht]
\centering \includegraphics[width=.7\columnwidth,angle=270]{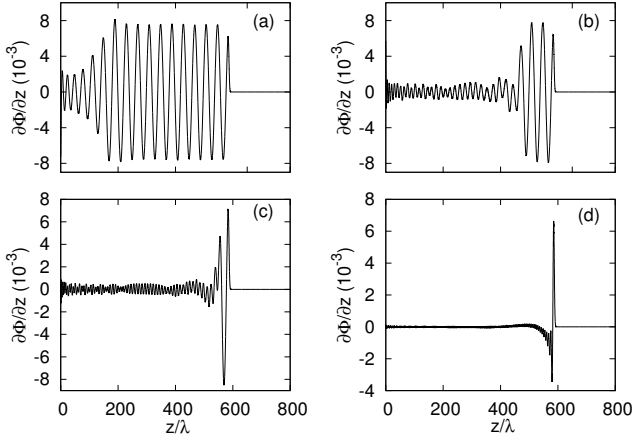}
\caption{Spatial profiles of the longitudinal wakefield at 600$\protect\tau$
in presence of external magnetic field with ratio $\protect\omega_c/\protect%
\omega_p =$ 0.5 (a), 2 (b), 4 (c) and 8 (d) respectively are presented by
numerically solving the Eqs. \protect\ref{laser-evolution} - \protect\ref%
{Wake-2}. Here we have converted the units of space and time in order to
compare with the PIC results.}
\label{WF_Ex_ana}
\end{figure}

\begin{figure}[ht]
\centering \includegraphics[width=.7\columnwidth,angle=270]{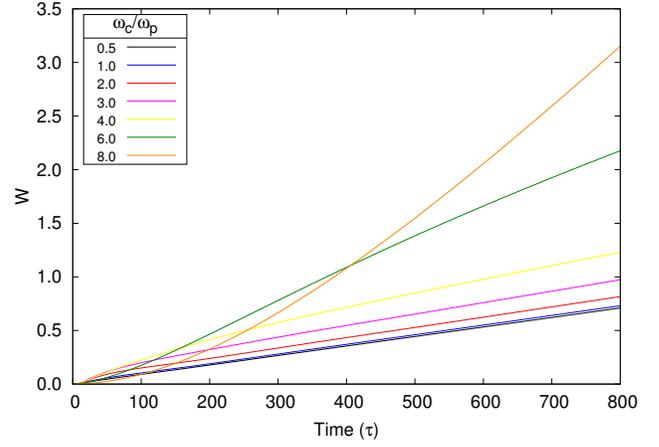}
\caption{Temporal evolution of the total energy of the wakefield (Eq. 
(\protect\ref{energy-density}) is integrated over space for each time step) for
different values of the external magnetic fields by numerically solving the
set of equations.}
\label{loss_SIMU}
\end{figure}

\begin{figure}[ht]
\centering \includegraphics[width=.7\columnwidth,angle=270]{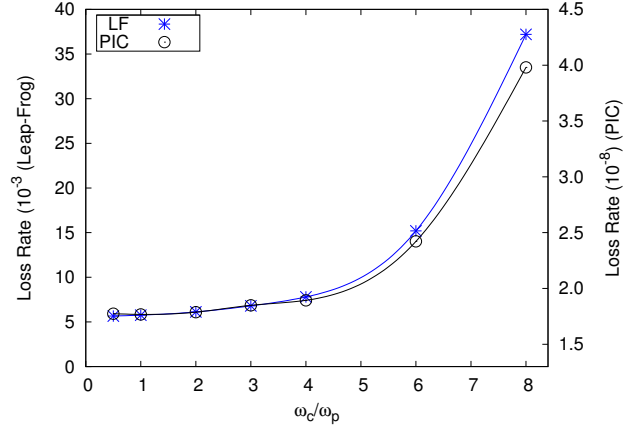}
\caption{Energy loss rate as a function of ratio $\protect\omega_c/\protect%
\omega_p$ is presented for PIC as well as analytical results.}
\label{lossRate}
\end{figure}

The numerical solutions of Eqs. (\ref{laser-evolution})- (\ref{Wake-2}) for
the longitudinal parts of the electric fields are shown in Fig. \ref{WF_Ex_ana} for
different ratios of $\omega _{c}/\omega _{p}$. Comparing these curves with
the results from the full simulations, shown in Fig. \ref{WF_Ex_PIC}, we see that the
agreement is excellent. 
%
%
Calculating the energy evolution of the wakefield (as based on the 
energy densities given in (\ref{energy-density}) and (\ref{energy-dens-driver})), 
which is displayed in Fig. \ref{loss_SIMU}, we furthermore get convincing agreement with the 
PIC-results shown in Fig. \ref{loss_PIC}. A more direct quantitative comparison of the energy 
loss rate is also shown in Fig. \ref{lossRate}, where the PIC-simulation and the
numerical results from (\ref{laser-evolution})--(\ref{Wake-2}) are simultaneously displayed.
We thus conclude that the wakefield generation can be seen as the
extra-ordinary mode driven by the ponderomotive force, in agreement with the
model (\ref{laser-evolution})- (\ref{Wake-2}). However, in contrast to an
assumption commonly made in connection with wakefield generation, the
evolution is not necessarily slow in a frame moving with the group velocity
of the driver. To be more specific - for a low to modest ratio of $\omega
_{c}/\omega _{p}$, the assumption of slow evolution in the co-moving frame
is adequate. However, starting roughly from the value $2\lesssim
\omega _{c}/\omega _{p}$ it is not appropriate to consider the
evolution as slow, which is reflected by the broad-band nature of the
wakefield spectrum (see Fig. \ref{FFT_Ez}). It should be noted that besides abandoning the 
slow evolution assumption when deriving the Eqs. (\ref{laser-evolution})- (\ref{Wake-2}) we have 
also refrained from the WKB-approximation of the high-frequency driver, 
that were made in Ref. \cite{PRE1998}. By contrast to the slow evolution 
assumption, the validity of the WKB-approximation for the
driver does not depend only on the ratio of $\omega _{c}/\omega _{p}$. 
Instead the parameter $kL$ $\ $(where $L$ is the pulse length, and $k$ the
high frequency wavenumber) naturally plays a major role. In Fig. \ref{FFT_Ez} we see 
that the wakefield frequency spectrum extends almost a factor 6 above 
the value $\omega _{p}$. For efficient wakefield generation typically 
$\omega _{p} L/c\sim 1$. Thus, in case $kL\gg 6$, the high-frequency part of the 
wakefield spectrum is still much lower than the frequency of the high-frequency 
driver, and a WKB-approximation for this variable would be valid, whereas
for $kL\sim 6$ it should be avoided due to the interaction with the turbulent 
wakefield spectrum which has a component of comparatively high frequency. 
The parameters used in our case, $kL=40$, lies somewhere in
the intermediate range, and therefore we have taken the safe approach 
and avoided the WKB-approximation for the high-frequency driver, although it
should clearly be applicable for a sufficiently large value of $ kL$, 
even for the highest values of $\omega _{c}/\omega _{p}$ that has been 
considered in our paper.

\section{Summary and Conclusion}

In the present paper we have re-considered 1D wakefield generation in
magnetized plasmas. Based on 1D PIC simulations we have deduced that
previous results need to be modified in the regime of large magnetic fields.
For moderate magnetic field strengths previous results are confirmed, and
the wakefield consists of an extra-ordinary wave mode with a dominating
frequency $\omega_p$ and wavenumber $\omega _{p}/c$. 
However, monitoring the dependence as a
function of the frequency ratio $\omega _{c}/\omega _{p}$, we find that
there is a gradual transition in the regime $2\lesssim \omega _{c}/\omega
_{p}\lesssim 4$ where the wakefield spectrum of the extra-ordinary mode
becomes broadband, and the loss rate of the high-frequency driver is
enhanced..

Although we have deduced that wakefield generation is governed by Eqs. (\ref%
{laser-evolution})- (\ref{Wake-2}), we have not so far discussed why the
assumption of a slow evolution in the co-moving frame must be abandoned for
a large value of $\omega _{c}/\omega _{p}$. However, an explanation of this
can be given in the form of a \textit{reductio ad absurdum }reasoning. Let
us assume that evolution of the extra-ordinary mode is indeed slow in the
co-moving frame. In that case, there must be a dominant wavenumber of the
wakefield corresponding to the property $v_{\phi wf }=v_{g}\approx c$, where 
$ v_{g}$ is the group velocity of the high-frequency driving pulse and 
$v_{\phi wf }$ is the phase velocity of the wakefield. However,
from the dispersion relation of the extra-ordinary mode we find that the
group velocity for this wavenumber scales according to $v_{g}\rightarrow c$
for $\omega _{c}/\omega _{p}\rightarrow \infty$. This behavior is to a
large extent reflected in Figs \ref{WF_Ex_PIC}, \ref{WF_Ez_PIC} and \ref{FFT_Ez}, where the dominant mode follows
the driver closely for the stronger values of the magnetic field. On the
other hand provided the wakefield should follow the driver closely (as it
must when $v_{g}\rightarrow c$) and then drop in amplitude suddenly, it
cannot be well represented by a small spectrum surrounding a leading
wave-number. Hence we must have broad-band wakefield spectrum for strong
magnetic fields, in which case the assumption of a slow evolution of the
wakefield will cease to be adequate.

To some extent the break-down of the slow evolution of the wakefield is
related to the initial conditions. After long interaction times the system
may gradually set up a wakefield with a dominating wave number, and at the
same time the slow evolution regime can be approached. However, the
stronger the magnetic field is, the longer time this will take. For a
limited size of the plasma region and a strong magnetic field, in practice the slow
evolution regime will never be reached. A practical consequence of this is
that the loss rate of the high-frequency driver will be much enhanced, due
to the excitation of the broad-band spectrum. Furthermore, the energy loss
will be accompanied by a frequency decrease and \ a de-acceleration of the
driver \cite{Mironov1990}.

So far only the basic consequences of a strong magnetic field has been
explored. A more complete study, e.g. involving the effects of varying the
self-nonlinearity and the long-time evolution, is a project for further
study.

\acknowledgments
This work is supported by the Baltic Foundation, the Swedish Research Council Contract \# 2007-4422 and the European Research Council Contract \# 204059-QPQV. This work is performed under the \emph{Light in 
Science and Technology} Strong Research Environment, Ume{\aa} University.


\begin{thebibliography}{99}
\bibitem{NOTE} The coordinate system used in LPIC++ is different. Here we have changed the
labels to make it consistent with rest of the paper.
\end{thebibliography}
\end{document}